\documentclass[twocolumn]{aastex62}
\usepackage[utf8]{inputenc}

\usepackage[nointegrals]{wasysym}
\usepackage{amsmath}
\usepackage{graphicx}
\usepackage{commath}
\usepackage{bm}
\usepackage{float}

\begin{document}

\title{SYNTHETIC LIGHT CURVES OF ACCRETION VARIABILITY IN T TAURI STARS}

\author[0000-0003-1639-510X]{Connor E. Robinson}
\author[0000-0001-9227-5949]{Catherine C. Espaillat}
\affiliation{ Department of Physics \& Astronomy, Boston University, 725 Commonwealth Avenue, Boston, MA 02215, USA}
\author[0000-0002-4856-7837]{James E. Owen}
\affiliation{Astrophysics Group, Department of Physics, Imperial College London, Prince Consort Rd, London SW7 2AZ, UK}

\shortauthors{Robinson et al.}
\shorttitle{Variability in T Tauri stars}
\email{connorr@bu.edu}

\begin{abstract}

Photometric observations of accreting, low-mass, pre-main-sequence stars (i.e., Classical T Tauri stars; CTTS) have revealed different categories of variability. Several of these classifications have been linked to changes in $\dot{M}$. To test how accretion variability conditions lead to different light-curve morphologies, we used 1D hydrodynamic simulations of accretion along a magnetic field line coupled with radiative transfer models and a simple treatment of rotation to generate synthetic light curves.  We adopted previously developed metrics in order to classify observations to facilitate comparisons between observations and our models.  We found that stellar mass, magnetic field geometry, corotation radius, inclination, and turbulence all play roles in producing the observed light curves and that no single parameter is entirely dominant in controlling the observed variability. While the periodic behavior of the light curve is most strongly affected by the inclination, it is also a function of the magnetic field geometry and inner disk turbulence. Objects with either pure dipole fields, strong aligned octupole components, or high turbulence in the inner disk all tend to display accretion bursts. Objects with anti-aligned octupole components or aligned, weaker octupole components tend to show light curves with slightly fewer bursts. We did not find clear monotonic trends between the stellar mass and empirical classification. This work establishes the groundwork for more detailed characterization of well-studied targets as more light curves of CTTS become available through missions such as the Transiting Exoplanet Survey Satellite (TESS).

\end{abstract}

\section{Introduction}

Variability has been a nearly ubiquitous feature of accreting, pre-main-sequence, low-mass stars known as Classical T Tauri stars (CTTS) since their discovery \citep{joy45, herbig62}, and has been observed on timescales ranging from minutes to decades \citep{siwak18}. One of the major sources of variability that occurs on timescales of hours to weeks is variability in the accretion rate onto the star \citep{herbst94}. These changes are most readily apparent in near-ultraviolet (NUV) and the optical. 

Our current understanding of how matter is accreted from the innermost region of the disk onto the stellar surface is through the process of magnetospheric accretion \citep[see][for a review]{hartmann16}. Magnetic fields on young stars are remarkably strong, with surface strengths reaching red up to several kiloGauss \citep{basri92, johnskrull99, valenti04, johnskrull07}, which is strong enough to disrupt the Keplerian motion of the disk and force flow along magnetic field lines red if the large-scale structure of the magnetic field is sufficiently globally ordered \citep{koenigl91, shu94}. The disrupted material falls along the magnetic field lines toward the star, reaching free fall speeds of 300--500 km/s. When this material collides with the stellar photosphere, it shocks and is heated to roughly $10^6$K \citep{calvet98}. As the material falls and cools, it thermally emits this energy primarily in the form of X-rays both downward toward the stellar surface and upward into the preshock region. In these regions, the high-energy radiation is reprocessed and emitted as far-ultraviolet (FUV), near-ultraviolet (NUV), and optical continuum emission, along with a plethora of emission lines \citep{gullbring98, ardila13}. 

Recent space-based long-timescale optical photometric monitoring campaigns from CoRoT \citep{cody14, stauffer14, stauffer15, stauffer16, mcginnis15} and K2 \citep{cody18} have been successful in classifying CTTS variability in several categories based on their light-curve symmetry and periodicity by-eye and subsequently using statistical metrics to separate objects by light-curve periodicity and symmetry ($Q$ and $M$, respectively). 
These classifications include bursters, quasiperiodic symmetric objects, purely stochastic objects, and purely periodic objects, whose variability has in part been attributed to changes in the accretion rate and/or to rotational effects from stable accretion hot spots \citep{stauffer14, venuti15}. Young stars that predominantly show decreases in observed flux are classified as quasiperiodic dippers and aperiodic dippers \citep[e.g.,][]{stauffer15, kesseli16, nagel19}.  Objects that do not fit into these categories are placed into the classes of long timescale, unclassifiable, non-variable, or eclipsing binary \citep[e.g.,][]{cody18}. 

Many works have described the accretion variability and the inner regions of protoplanetary disks under the constraint of magnetically controlled accretion. Steady-state solutions of 1D flow along the accretion column have been studied analytically by several authors \citep[e.g.,][]{hartmann94, li96, koldoba02}. Additional work has been done with 3D MHD simulations of the inner regions of the disk \citep[e.g.,][]{romanova03, romanova04, romanova08, long11, romanova11, kurosawa13, romanova13}. These works identify numerous sources of variability including rotational modulation and Rayleigh--Taylor instabilities. 
One example of a semi-analytic approach is the work of \citet{adams12}, which described steady-state 1D solutions within a coordinate system that follows along a corotating magnetic field line connecting the star to the disk. That work was expanded upon by \citet{robinson17} using simulations that demonstrated the formation of traveling shocks along the magnetospheric accretion column under smoothly varying disk input conditions. These shocks cause rapid sawtooth-like changes in the accretion rate that qualitatively resemble the variability seen in objects under the burster classification. However, the variability in those simulations is induced through density perturbations controlled by an overly simplified sinusoidal driving function. Additionally, those simulations were not used to make direct predictions of the accretion events to compare against observations.

Here, we combine the 1D hydrodynamic fluid simulations of the accretion column of \citet{robinson17} and radiative transfer accretion shock models \citep{calvet98, robinson19} to produce the expected excess from accretion under a variety of physical conditions in the inner disk. This excess is then characterized using the same metrics that have been applied in long-timescale optical monitoring studies of CTTS. In \S 2, we present our models and describe the process of obtaining the metrics. In \S 3, we demonstrate how different physical parameters affect light curves. In \S 4, we discuss our findings within the context of other studies and possible future work. We conclude with a summary of our primary findings in \S 5. 

\section{Methods}

\subsection{Overview}

To create synthetic light curves, we couple three models together. These include a 1D hydrodynamic (HD) simulation of the accretion column; nonlocal thermodynamic equilibrium (nLTE) accretion shock models of the preshock, postshock, and heated photosphere; and a geometric rotational modulation model. 
We run the 1D HD simulations of the accretion column to obtain the density and velocity of the flow near the surface of the star. The emission arising from the shock is then found with the nLTE accretion shock model by using the conditions of the flow from the HD simulations as input. That emission is then scaled by the fraction of the accretion hot spot that is visible based on the rotational phase, inclination, and spot latitude (i.e., a filling factor). 
We note that no feedback is included in the HD simulations from the nLTE shock models and that the solutions from the nLTE models are obtained in a time-independent manner.  If the radiative timescales are longer than the accretion variability, then this approach may break down. However, we note that the hydrogen recombination timescales within the pre- and post-shock regions are likely several orders of magnitude shorter than the variability considered here. We finish this section by describing the metrics and relevant variability classes that have been developed to describe  observed light curves of CTTS that we then apply to the light curves from our models. Definitions for commonly reoccurring symbols used in this work are listed in Table \ref{tab:vartab}.

\begin{deluxetable}{cc}
\centering
\tablecolumns{2}
\tablewidth{\columnwidth} 
\tablecaption{\label{tab:vartab} Reoccurring Symbols used in This Work} 
\tablehead{Symbol & Definition}
\startdata
\sidehead{\textbf{Coordinate Systems \& Magnetic Field}}
$\theta$ & Polar angle  \\
$\phi$ & Azimuthal angle \\
$r$ & Radial coordinate \\
$p$ & Distance from origin along field line  \\
$q$ & Coordinate that selects the field line \\
$\xi$ & Dimensionless radial coordinate \\
$B_{dip}$& Polar dipole magnetic field strength \\
$B_{oct}$ & Polar octupole magnetic field strength \\
$\Gamma$ & Ratio of $B_{oct}$ to $B_{dip}$ \\
\sidehead{\textbf{Star, Disk, \& Fluid}}
$R_{co}$ & Corotation radius \\
$R_{T}$ & Truncation radius, fixed to $0.5 R_{co}$\\
$i$ & Inclination \\
$A$ & Turbulent perturbation amplitude \\
$n_0$ & Number density at the disk\\
$F$ & Kinematic energy flux of the flow \\
\sidehead{\textbf{Rotational Modulation}}
$f$ & Accretion filling factor for the visible stellar surface \\
$\alpha$ & Longitudinal spot width \\
$\gamma_U$ & Upper bound of hot spot \\
$\gamma_L$ & Lower bound of hot spot \\
\sidehead{\textbf{Light-curve Metrics}}
$Q$ & Periodicity metric \\
$M$ & Symmetry metric \\
$\sigma$ & Injected white noise  \\
GP & Gaussian processes  \\
& 
\enddata
\end{deluxetable}

\subsection{Hydrodynamic Simulations}

To model the accretion column connecting the star to the inner edge of the disk, we turn to the 1D HD fluid simulations of \citet{robinson17}. These simulations follow along the magnetic field line that connects the inner gas disk to the star, following the curvilinear coordinate system developed by \citet{adams12}. These simulations operate under several simplifying assumptions that enable the use of a 1D approach. The assumptions and the implication for each assumptions are described below.

\subsubsection{Assumptions}

The magnetic field is assumed to be current free, and therefore curl free, and thus can be expressed as the gradient of a scalar field. We take advantage of this assumption by constructing a set of orthogonal curvilinear coordinates to describe the magnetic field lines. The coordinate $p$ describes the distance from the origin (which is at the center of the star) along a field line, while the orthogonal coordinate $q$ describes which field line is being traced. The third coordinate, $\phi$, describes the position in the azimuthal direction. 

The magnetic field is assumed to be strong enough to force flow solely along the field line, which allows us to work in a 1D space described solely by $p$ under a fixed $q$ and $\phi$. Thus, many complexities that can arise in higher dimensions (e.g., several types of instabilities) are suppressed. Compression from converging field lines is included through curvilinear scale factors. The assumed magnetic field structure is that of a rotation-axis-aligned multipolar field. While measurements of the magnetic field structure of some CTTS have shown evidence for complex toroidal fields \citep{donati11c}, many CTTS can be well described by composite fields consisting of dipole and octupole components \citep[e.g., ][]{donati10a, donati11b}. Following \citet{gregory10, gregory11, adams12}, the ratio of the polar field strengths of the dipole and octupolar magnetic field components is parameterized by $\Gamma \equiv \frac{B_{oct}}{B_{dip}}$ with the magnetic field described by
\begin{equation}
\begin{split}
    \mathbf{B} = \frac{B_{oct}}{2} \xi^{-5} \Big[(5\cos^2\theta - 3)\cos\theta \, \hat{r} 
     + \frac{3}{4}(5\cos^2\theta - 1)\sin \theta \, \hat{\theta}\Big] + \\
     \frac{B_{dip}}{2} \xi^{-3} (2\cos\theta \, \hat{r} + \sin\theta \, \hat{\theta}),
\end{split}
\end{equation}
where $\xi \equiv \frac{r}{R_\star}$. In the aligned case where the dipole and octupole moments are parallel, $\Gamma$ is positive. In the anti-aligned case where either of the moments are flipped, $\Gamma$ is negative and the positive pole of the dipole field coincides with the main negative pole of the octupole field. In both cases, we make the assumption that the magnetic moments of both the dipole and octupole fields are perfectly aligned with the rotational axis of the star.

Our final assumption is that the solution for flow along the magnetic field line at the corotation radius (defined here as $R_{co}$) is an appropriate approximation of the entire column. By definition, the magnetic footprints at the star and the disk have the same angular velocity for the magnetic field line that crosses the disk at the corotation point. This eliminates the possibility of shear flows at either boundary. Following \citet{getman08a, getman08b, gregory08}, we make the assumption that the inner edge of the region where the magnetic field line draws material from the disk (i.e., the truncation radius) is $R_{T} = 0.5 R_{co}$ and the outer edge is $R_{co}$. Under the construction of our model, these two radii, along with $\Gamma$, set the upper and lower latitudes of the accretion hot spot, $\gamma_U$ and $\gamma_L$.

A more complete description of the simulations is shown in \citet{robinson17}, and more discussion on the curvilinear coordinate system can be found in \citet{adams12}. The top panel of Figure \ref{fig:schematic} shows a schematic of the system under several magnetic field constructions.

\subsubsection{1D HD Simulation Scheme and Boundary Conditions}

A ZEUS-style upwind finite difference scheme was used to solve the 1D fluid equations \citep{stone92}. Under this scheme, the density, velocity, and internal energy of the flow are solved on a staggered mesh. In this work, we restrict the flow along the accretion column to be isothermal at a temperature of 10,000K from UV heating, which sets the sound speed in the simulation. Simulations with non-isothermal equations of state demonstrated variability with similar characteristics as isothermal simulations in the work of \citet{robinson17}. The internal structure of the solver remains unchanged between this work and \citet{robinson17}, with the exception of the outer boundary conditions. 

\begin{figure}
    \centering
    \includegraphics[width = .97\linewidth]{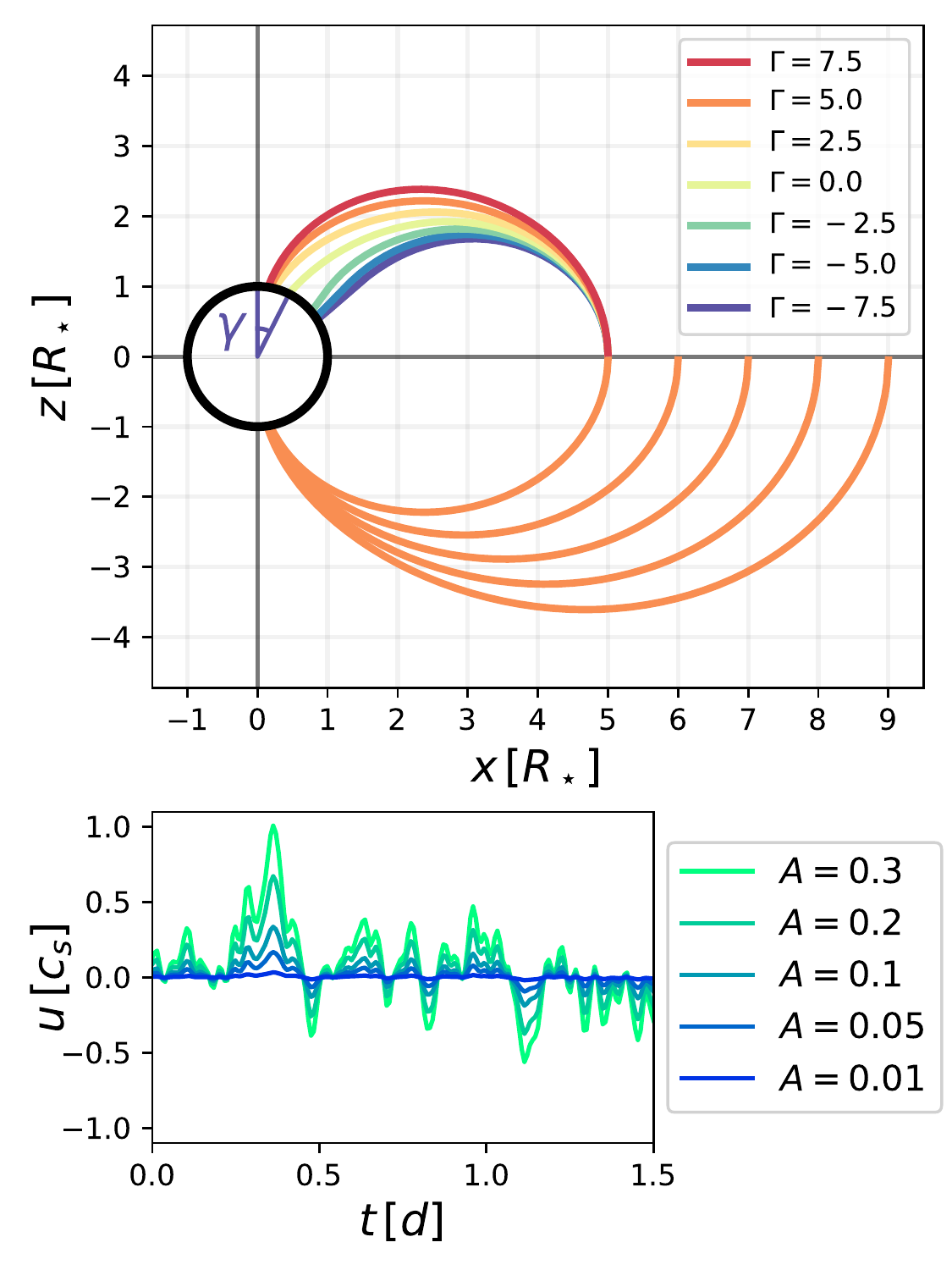}
    \caption{\textit{Top:} magnetic field line geometry for several values of the ratio between the polar strength of the octupole and dipole components, $\Gamma$, and corotation radius, $R_{co}$. Changes in $\gamma_{U}$ and $\gamma_L$, the angles between the upper and lower bound of the hot spot and the rotation axis, respectively, affect the degree of rotational modulation (see Figure \ \ref{fig:spotplot}). After normalizing to the area of the magnetic footprint at the disk, the effective cross-sectional area near the star decreases with increasing $\Gamma$ and $R_{co}$. This schematic is to scale, and the axes are in Cartesian coordinates in units of stellar radii. \textit{Bottom:} A 1.5 d sample of the turbulent driving function with several Mach number rms amplitudes, $A$. }
    \label{fig:schematic}
\end{figure}

The new boundary condition that has been implemented at the outer boundary mimics Kolmogorov turbulence \citep[see][]{kolmogorov41} in the inner disk by using a velocity driving function. The upper bound on the turnover frequency is set by a cutoff on events happening on timescales shorter than 30 minutes. The lower frequency bound is set by the travel time of a sound wave around the circumference of the inner disk of our base model (described in \S 3.2). A turbulence spectrum with these bounds is then used to generate velocity driving functions composed of sine curves with random phase offsets. We do not restrict outflow from the system, and as such, these boundary conditions result in solutions and values of $\dot{M}$ similar to those under steady-state boundary conditions with the addition of stochastic perturbations. The system is allowed to evolve for 18 days ($\sim14$ sound crossing times) to allow transients from initialization to disappear before the simulation is sampled. An example of one of the driving functions that mimics turbulence is shown in the bottom panel of Figure \ref{fig:schematic}.

\subsection{Accretion Shock Model}

The nLTE accretion shock models of \citet{calvet98}, which were recently adapted by \citet{robinson19} to use a recent release of Cloudy \citep{ferland17}, were used to generate the emission from accretion. This model consists of a 1D plane-parallel accretion column with three parts: the preshock, the post-shock, and the heated photosphere. The post-shock is hot ($\sim10^6$ K) due to the kinetic energy of the flow being converted into thermal energy in the shock. As the material cools and falls, it re-radiates that energy (primarily as X-rays) into the preshock region and the underlying photosphere. The combined emission from these two regions forms the outgoing flux. Previous work has demonstrated the need to include accretion columns with different densities to reproduce observations \citep{ingleby13, ingleby15, robinson19}. 

Here, we slightly modify the \citet{robinson19} models by using the density and velocity information from the 1D fluid simulations as the preshock conditions rather than assuming that the flow is at the freefall velocity.  In general, this causes the accretion column to be denser and slower than previous models at the same kinetic energy flux. This in turn causes the post-shock region to narrow (typically by factor of $\sim0.7$) since the region is assumed to be optically thin and the velocity at the top of the region depends on the preshock velocity via the Rankine--Hugoniot strong shock conditions. Apart from this change, the accretion shock model remains unchanged from the version presented in \citet{robinson19}.

\subsection{Rotational Modulation Model \label{ss:rotation}}
\begin{figure}
    \centering
    \includegraphics[width = \linewidth]{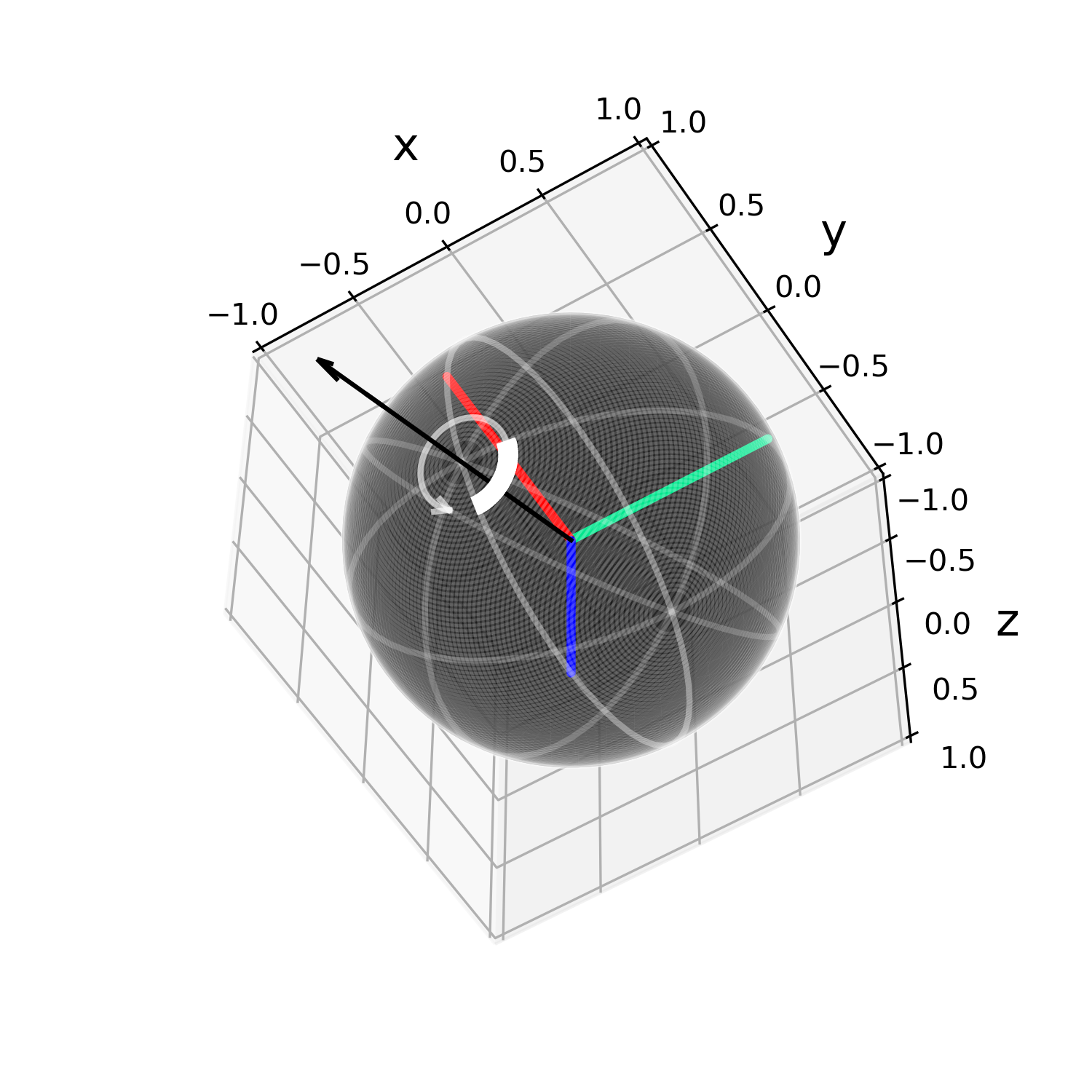}
    \caption{Schematic showing the the geometry of the rotational modulation model. The green, red, and blue lines represent the $x$, $y$, and $z$ axes, respectively. The observer is located along the $z$-axis. The black arrow represents the rotation axis of the star, shown here at an inclination of $60^\circ$. The white, equilatitudinal, rectangular strip is the accretion hot spot for a model with $\alpha = 90^\circ$, $\Gamma = 5$, and $R_{co} = 5 \, \mathrm{R_{\star}}$ (i.e., the base model, see \S \ref{ss:basemodel}). The white arrow shows the path of the hot spot as the star rotates. }
    \label{fig:sphereplot}
\end{figure}

The emission from the accretion shock model is scaled by a filling factor $f$, which represents the fraction of the visible stellar disk covered by accretion shocks. To estimate $f$, we have modeled the accretion hot spots on the surface of the star as an equilatitudinal, roughly rectangular, strip with angular length defined as $\alpha$ and angular width defined by the magnetic field. The filling factors are calculated by numerically integrating the projected area of the strip along the line of sight. An example of the geometry of this model is shown in Figure \ref{fig:sphereplot}.

The lower and upper bounds ($\gamma_{L}$ and $\gamma_{U}$) of the strip are set by the magnetic field geometry by following the magnetic field lines that intercept the disk at 0.5 $R_{co}$ and $1.0 R_{co}$, respectively. These bounds can be calculated under our assumption of flow solely along the rotation-axis-aligned dipole + octupole field.
Following \citet{adams12}, the general expression for the coordinate $q$, which is constant along field lines, is given by
\begin{equation}
q = \frac{1}{4}\xi^{-3} \Gamma (5\cos^2\theta - 1)\sin^2\theta + \xi^{-1}\sin^{2}\theta ~~.
\label{generalq}
\end{equation}
The value of the coordinate $q$ that describes a field line that crosses the midplane at a given radius can be written
\begin{equation}
    q = \frac{1}{\xi}\left(1 - \frac{\Gamma}{4 \xi^2}\right) ~~.
\label{q_cross}
\end{equation}
Because $q$ is a fixed quantity along field lines, Eq. \ \ref{generalq} can be solved for $\theta$ at the stellar surface with the value of $q$ from Eq. \ref{q_cross}. Under our construction, $\theta$ is equivalent to $\gamma_L$ and $\gamma_U$ for the inner and outer bounds of the disk--star interaction region, respectively.
The length of the spot, $\alpha$, and the viewing inclination, $i$, remain as free parameters.

\begin{figure*}
    \centering
    \includegraphics[width = \linewidth]{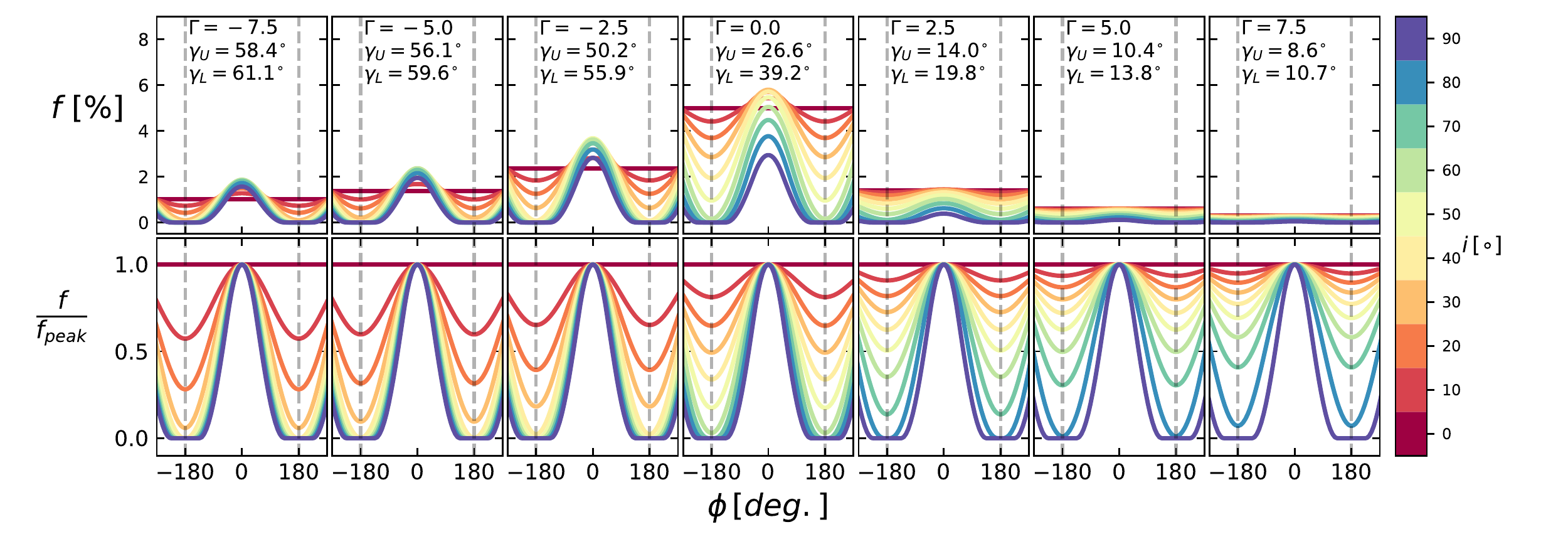}
    \caption{ \textit{Top row:} filling factor (expressed as a percentage) as a function of rotational phase, inclination, and magnetic field geometry for a single hot spot modeled as an equilatitudinal strip. The examples shown here have a fixed angular strip length of $\alpha = 90^\circ$ and $R_{co} = 5 \, \mathrm{R_\star}$. As $\Gamma$ increases, the magnetic footprint of the corotating field line on the star approaches the rotation axis.  The upper and lower bounds of the strip, as measured from the rotation axis, are $\gamma_U$ and $\gamma_L$, respectively. Due to the construction of the magnetic field, $\gamma_U$ and $\gamma_L$ are also anticorrelated with the inner disk radius. \textit{Bottom row:} filling factors as a function of $\phi$, normalized to their peak value to show the relative change between their minimum and maximum values.}
    \label{fig:spotplot}
\end{figure*}

The projected area of the strip as a function of phase is approximated by breaking the strip into several equilatitudinal rows of triangular cells and numerically integrating the projected area of each cell along the line of sight. Regions behind the disk of the star are discarded when summing the contribution from each cell. The numbers of rows and cells were selected to be large enough to avoid potential numerical issues associated with discretization. Examples of the filling factor as a function of $\phi$, $\Gamma$, and $i$ with $\alpha = 90^\circ$ are shown in Figure \ref{fig:spotplot}.

\subsection{Statistical Metrics $Q$ and $M$}

The metrics $Q$ and $M$ have recently been developed to classify light curves obtained using K2 and CoRoT \citep[see ][]{cody14, cody18}. Here, we apply these metrics to light curves produced using our simulations and models to provide physical insight to previous classifications of observations. The metric $Q$ describes the degree of periodicity in a given light curve, while $M$ quantifies the symmetry of the light curve around the mean to determine if the light curve in general shows bursts, dips, or no preference. 

\subsubsection{Determining $Q$}

\begin{figure*}
    \centering
    \includegraphics[width = 0.9\linewidth]{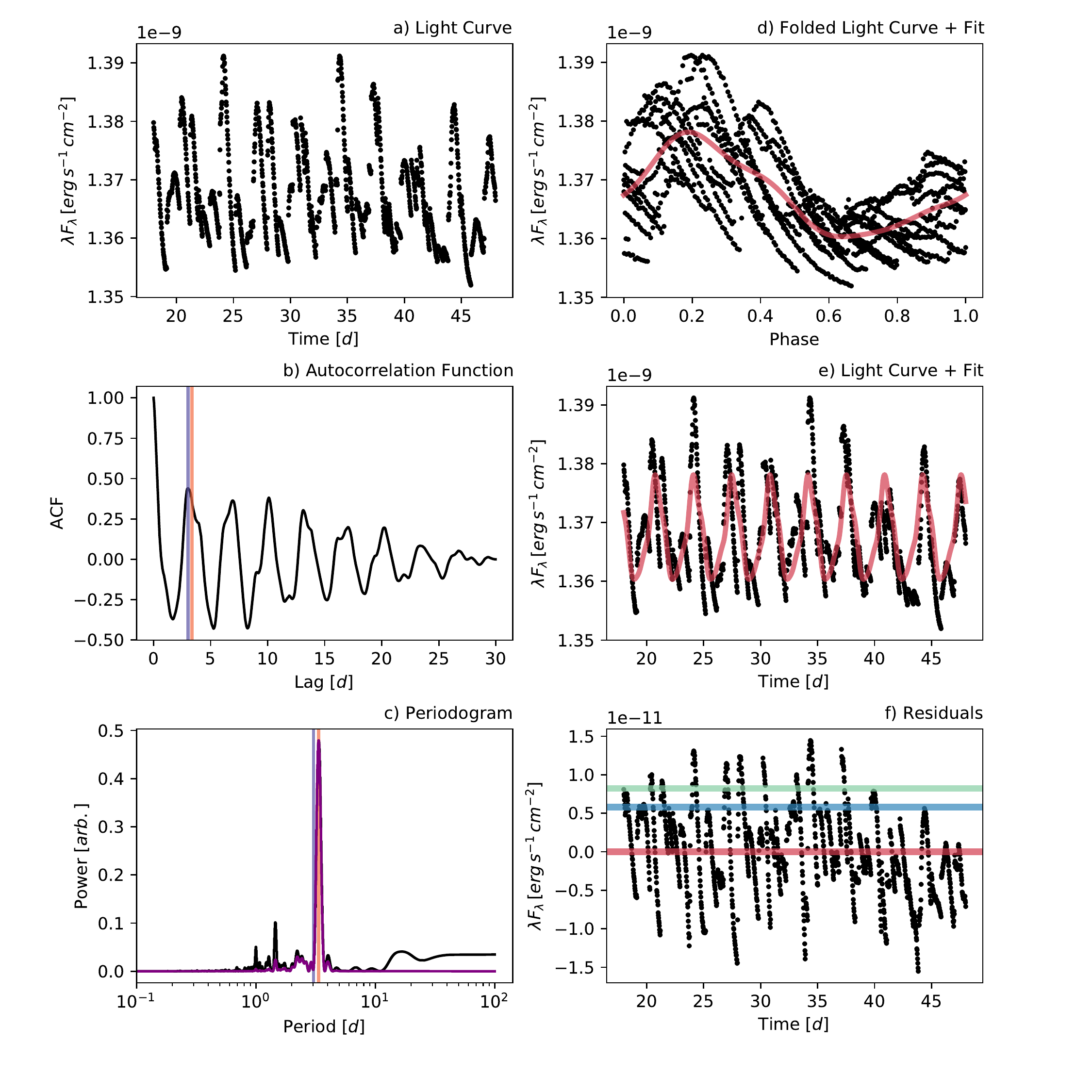}
    \caption{Illustration of the steps to obtain the metric $Q$, which measures the periodicity of the light curve (see \S 2.5.1 for details on the process).  This light curve is the base model described in \S3.2 viewed from an inclination of $60^\circ$. Panel (a) shows the raw light curve. Panel (b) is the autocorrelation function (ACF) with the period as measured by the ACF and the Lomb--Scargle periodogram marked by purple and orange lines, respectively. Panel (c) shows the Lomb--Scargle periodogram with the periods marked in the same fashion. We show the periodogram weighted by a Gaussian whose parameters are derived from the ACF in purple. Panel (d) shows the folded light curve with the GP solution shown in red. Panel (e) shows the raw light curve with the GP solution overlaid in red. Panel (f) shows the residuals after removing the GP solution. The horizontal red, blue, and green lines show the mean, $\mbox{rms}_{\mbox{resid}}$, and $\mbox{rms}_{\mbox{raw}}$, respectively. $Q$ in this example is 0.51.}
    
    \label{fig:gettingQ}
\end{figure*}

$Q$ is calculated following the steps outlined in \citet{cody14}, with a few modifications. The first step is identifying the rotation period. While the rotation period is known a priori in our simulations, we wish to conduct our analysis without including this prior knowledge. Additionally, we have injected a small amount of white noise into our synthetic light curves with amplitudes of $\sigma = 0.01\%$ of the mean value. Such a small value was chosen to avoid white noise from interfering with small amplitude variations since we are attempting to identify trends with free parameters rather than reproducing observations exactly. The process to measure $Q$ is described below and an example of the process is depicted visually in its entirety in Figure  \ref{fig:gettingQ}. The unmodified light curve is shown in Figure \ref{fig:gettingQ}(a).

The rotation period is estimated using two approaches: the autocorrelation function (ACF; Fig.\ \ref{fig:gettingQ}(b)) and a Lomb--Scargle periodogram \citep{lomb76,scargle82} (Figure \ref{fig:gettingQ}(c)). 
The synthetic light curve is resampled to a linear grid when calculating the ACF and is smoothed using a Savitzky--Golay filter with a window size of three points to help avoid issues with local minima \citep{savitzky64}. 
The shortest period local maximum of the smoothed ACF that is 0.05 larger than than the surrounding local minima is identified. In the method of \citet{cody14}, the largest peak in the Lomb--Scargle periodogram that is within $\pm20\%$ of the previously identified period in the ACF is chosen as the rotation period of the star. However, we found that this method occasionally selected harmonics rather than the true period of our models, so we deviate from the approach of those authors slightly. Instead, we multiply the periodogram by a Gaussian centered on the selected peak of the ACF with a width of $30\%$ of the period from the ACF, and then select the peak from this weighted periodogram. We found that this method was effective at recovering the true period of our models even in the cases where the selected peak of the ACF was a harmonic.

The periodic behavior of the light curve was then identified by horizontally stacking three copies of the phase-folded data and modeling the trend using Gaussian processes (GP), as shown by  Figures \ref{fig:gettingQ}(d) and (e). By stacking the phased light curve in this manner and making predictions with GP for the central copy of the stacked curves, we avoid edge effects and recover a smooth, periodic, nonparametric curve that can be modified to smooth over features with different timescales. The covariance function selected for this approach is a Gaussian with a characteristic covariance length of 0.3 in phase space. The predicted values from this GP approach for the central copy of our stacked phased light curve are then subtracted from the original light curve, as shown by Figure \ref{fig:gettingQ}f to eliminate periodic components. Using the results from this analysis, $Q$ is defined by
\begin{equation}
    Q = \frac{(\mbox{rms}_{\mbox{resid}}^2 - \sigma^2)}{(\mbox{rms}_{\mbox{raw}}^2 - \sigma^2)} ~~,
\end{equation}
where $\mbox{rms}_{\mbox{resid}}$ is the rms value of the residuals after subtracting the periodic behavior, and $\mbox{rms}_{\mbox{raw}}$ is the rms of the original light curve. $Q$ is smaller for curves with stronger periodic components and larger for more stochastic light curves.

\subsubsection{Determining $M$}

\begin{figure*}
    \centering
    \includegraphics[width = 0.9\linewidth]{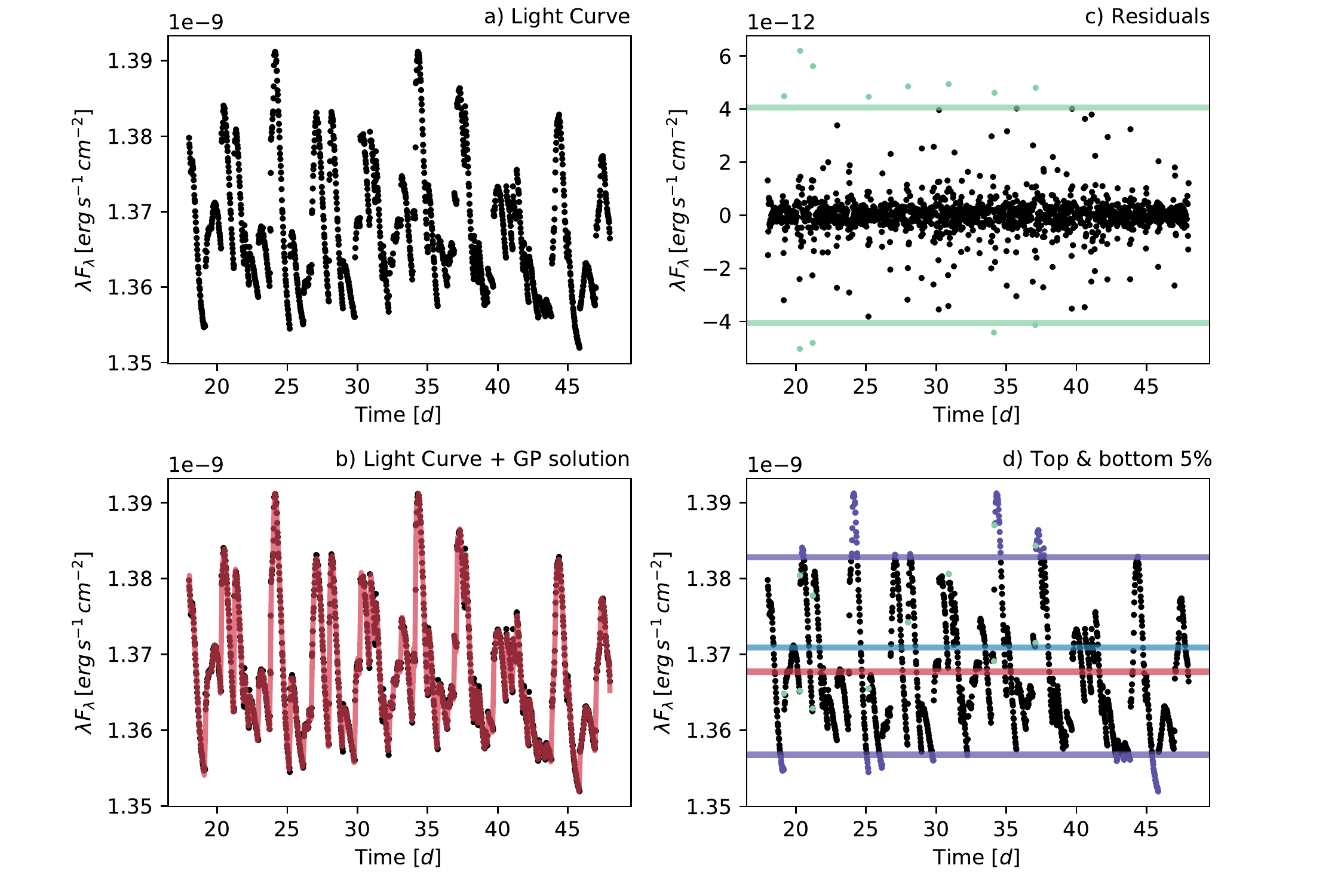}
    \caption{Illustration of the steps to obtain the metric $M$, which measures the symmetry of the light curve around the mean (see \S2.5.2). This light curve is produced using the base model described in \S3.2 with an inclination of $60^\circ$. Panel (a) shows the raw light curve. Panel (b) shows the light curve with the GP solution overlaid in red. Panel (c) shows the residuals after subtracting this fit. Points above the $5\sigma$ green lines are considered outliers. Panel (d) identifies the top and bottom 5\% of photometry points in purple, with the mean, $\langle d_{5\%}\rangle$, shown by the blue line. The median of all of the data, $d_{med}$,  is shown by the red line. $M$ in this example is -0.39.}
    \label{fig:gettingM}
\end{figure*}

Again following \citet{cody14} and \citet{cody18}, the first step for calculating $M$ is to remove outliers. This is accomplished by subtracting a smoothed version of our light curve from the raw light curve (Figures \ref{fig:gettingM}(a) and (b)). We again employ GP with a Gaussian covariance function to produce a smooth version of our light curve that is subtracted from the original curve (Figure \ref{fig:gettingM}(c)) in a flexible manner. The covariance length in this case was selected to be 2 hr. Outliers were identified as points in the residuals whose absolute values were greater than five times the rms. After clipping these outliers, the mean of the top and bottom $5\%$ of points from the original curve is found, $\langle d_{5\%} \rangle$ (Figure \ref{fig:gettingM}(d)). Using this mean, the median of the entire clipped light curve, $d_{med}$, and the rms of the clipped light curve, $\sigma_d$, the metric $M$ is defined 
\begin{equation}
    M = - \frac{\langle d_{5\%} \rangle - d_{med}}{\sigma_d} ~~.
\end{equation}
Here a negative sign is included since the original metric was constructed for use with magnitudes in \citet{cody14}. More negative values of $M$ indicate that the light curve features bursting behavior, while positive values of $M$ suggest preferential dimming. Values of $M$ closer to zero indicate symmetric variability.  

\subsection{Variability Classes}

The bivariate plot of $Q$ versus $M$ can be broken into rough regions that display different forms of variability \citep[e.g.,][]{cody14, cody18} based on visual classification. The definitions for the regions that we adopt \citep[from][]{cody18} are:
\begin{itemize}
    \item \textit{Burster} \\ $M < -0.25$ 
    \item \textit{Purely periodic symmetric} \\ $Q < 0.15$ and $-0.25 < M < 0.25$
    \item \textit{Quasiperiodic symmetric} \\ $0.15 < Q < 0.85$ and $-0.25 < M < 0.25$
    \item  \textit{Purely stochastic} \\ $Q > 0.85$ and $-0.25 < M < 0.25$ 
    \item \textit{Quasiperiodic dipper} \\ $0.15 < Q < 0.85$ and $M > 0.25$
    \item \textit{Aperiodic dipper} \\ $Q > 0.85$ and $M > 0.25$~~.
\end{itemize}
In addition to these categories, which can be sorted by their position in $Q$ and $M$ space, \citet{cody18} and \citet{cody14} include classifications by-eye for long-timescale, unclassifiable, non-variable, and eclipsing binaries. Here we primarily focus on the types of variability that are thought to be caused in part by accretion-related phenomena (burster, purely periodic symmetric, quasiperiodic symmetric, and purely stochastic).

\section{Results}

After linking the HD simulation, the nLTE shock model, and the model of rotational modulation, we explored which parameters had an effect on the observed variability through the $Q$ and $M$ metrics. Below we discuss each of the parameters and their effect on the light curve.

\subsection{Free Parameters}

The relevant parameters that can be readily changed in our models are as follows: the polar strength of the octupole relative to the dipole, $\Gamma$, which sets the magnetic field geometry (see Figure \ref{fig:schematic}); the corotation radius, $R_{co}$; the rms of the turbulent driving function, $A$; the stellar parameters, $M_\star$, $R_\star$, and $T_{\mathrm{eff}}$; the density of the accretion column at the disk, $n_{0}$; the inclination, $i$, of the star; the upper and lower latitudinal bounds of the hot spot strip (measured from the rotational axis), $\gamma_U$ and $\gamma_L$; the angular length of the hot spot strip, $\alpha$; the rotational period; and the filter bandpass. 

While the number of free parameters initially appears daunting, several of them can be eliminated or combined into a single variable. The three stellar parameters, $M_\star$, $R_\star$ and $T_{\mathrm{eff}}$ can be reduced to the single parameter of $M_\star$ if an age is assumed using isochrones. We used the MESA Isochrones and Stellar Tracks (MIST) with solar metallicity at an age of 2 Myr \citep{choi16, dotter16}. As noted in \S\ref{ss:rotation}, under our assumption of a multipolar field and an assumption on the inner/outer radius from which to draw material from the disk, we can solve for the spot latitude, which eliminates $\gamma_U$ and $\gamma_L$ as free parameters. The fluid equations can be solved in a dimensionless manner, which means that changing $n_0$ does not affect the flow structure along the column. While the value of $n_0$ affects the output spectra, $\dot{M}$, and magnetic field strength, changing it will not strongly affect the measured $Q$ and $M$ with the small value of $\sigma$ that we have adopted. Here, we fix $n_0$ to a single value of $10^{13}$~cm$^{-3}$, which results in kinetic energy fluxes between roughly $10^{10}$ and $10^{12} \, \mathrm{erg \, s^{-1} \, cm^{-2}}$ which is typical for young stars \citep{ingleby13, ingleby15, robinson19}. The rotational period of the star can be determined from the specified parameters above using Kepler's third law for circular orbits at the $R_{co}$.
Unless stated otherwise, the light curves are generated using the Johnson V filter \citep{johnson53}. From test runs of the model, we found that $Q$ and $M$ were generally only weakly dependent on the length of the hot spot, $\alpha$. Given this, we fix $\alpha = 90^\circ$, which for a pure dipole field results in a total surface coverage fraction of about $1.5\%$ (or $0.15\%$ for $\Gamma = 5$) for a single hot spot.

\subsection{Base Model \label{ss:basemodel}}

\begin{figure}
    \centering
    \includegraphics[width = .95\linewidth]{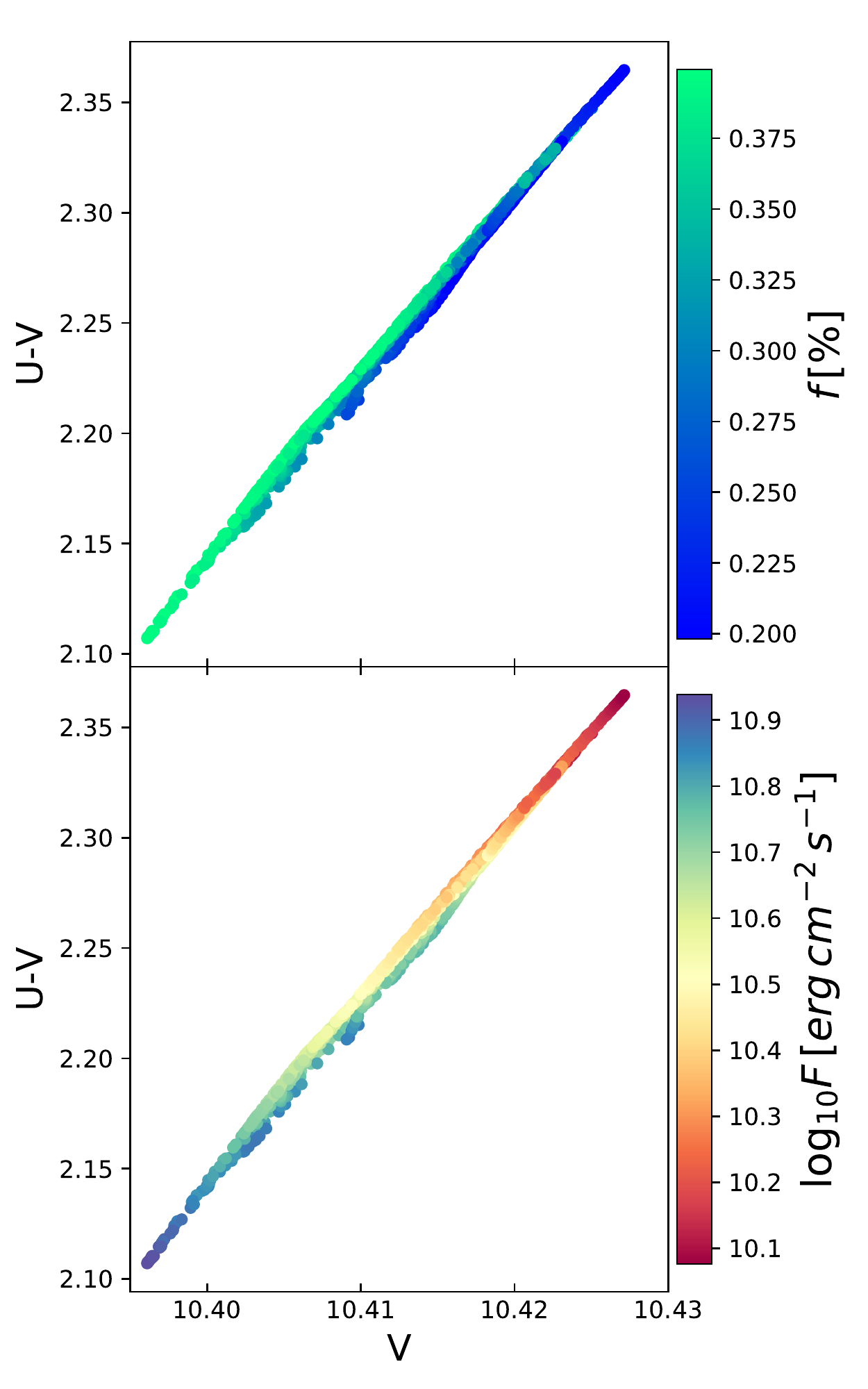}
    \caption{ $U-V$ vs. $V$ diagrams created using the timesteps from the base model at an inclination of $60^\circ$. The top panel is colored by the filling factor, $f$ (expressed as a percentage), while the bottom panel is colored by the kinetic energy flux, $F$. We find a gradient along the narrow diagonal locus of points for both of these quantities, demonstrating the degeneracy between both parameters at these wavelengths.}
    \label{fig:U-V}
\end{figure}

After the above simplifications, five relevant free parameters remain: $\Gamma$, $R_{co}$, $A$, $M_\star$, and $i$. To explore this parameter space, we first construct a base model with parameters thought to be somewhat typical of young stars. The base model has $\Gamma = 5.0$, $R_{co} = 5.0 \, \mathrm{R_{\star}}$, $M_{\star} = 1.0 \, \mathrm{M_\odot}$,  and a period of 3.3 days. Ten inclinations were evenly sampled between $0^\circ$ and $90^\circ$. Figure \ref{fig:U-V} shows the $U-V$ color against $V$ for this base model viewed at an inclination of $60^\circ$ for synthetic observations lasting 30 days. Fig \ref{fig:U-V} also demonstrates the degeneracy between the filling factor, $f$, and the kinetic energy flux of the flow at the shock, $F$, which can be written as
\begin{equation}
F = \frac{1}{2}\rho v^3
\end{equation}
where $\rho$ and $v$ are taken from the HD simulation at the stellar surface.

Our base model was then run three times using the turbulent driving functions described above, each with a different random seed value. $Q$ and $M$ were then calculated for each of these models. As a rough estimate of the uncertainty in $Q$ and $M$, we adopt the mean of the standard deviations between the three realizations at each inclination. The process of determining $Q$ and $M$ for one of the light curves from a model realization at $i = 60^\circ$ is illustrated in Figures \ref{fig:gettingQ} and \ref{fig:gettingM}, respectively. 
Typical, instantaneous $\dot{M}$ values for the selected base parameters with a single spot are on the order of $3 \times 10^{-10} \, \mathrm{M_\odot \, yr^{-1}}$. With these accretion rates and the approximation that the $R_T$ is set by the dipole field component, by inverting the typical calculation for the $R_T$ \citep[e.g., the derivation in ][]{hartmann16} we find that typical values of $B_{dip}$ for the base model are around $60$ G with a total field strength of $|B| \sim 360$ G near the poles. This value of $B_{dip}$ is low compared to those found many spectropolarimetric studies \citep[e.g.,][]{donati08b, donati10b, donati11a, donati11b}, but we note that many of the values of $B_{dip}$ for the model variants discussed in the following section are closer to the observed values.

\subsection{Variants on the Base Model}

\begin{figure*}
    \centering
    \includegraphics[width = 0.9\linewidth]{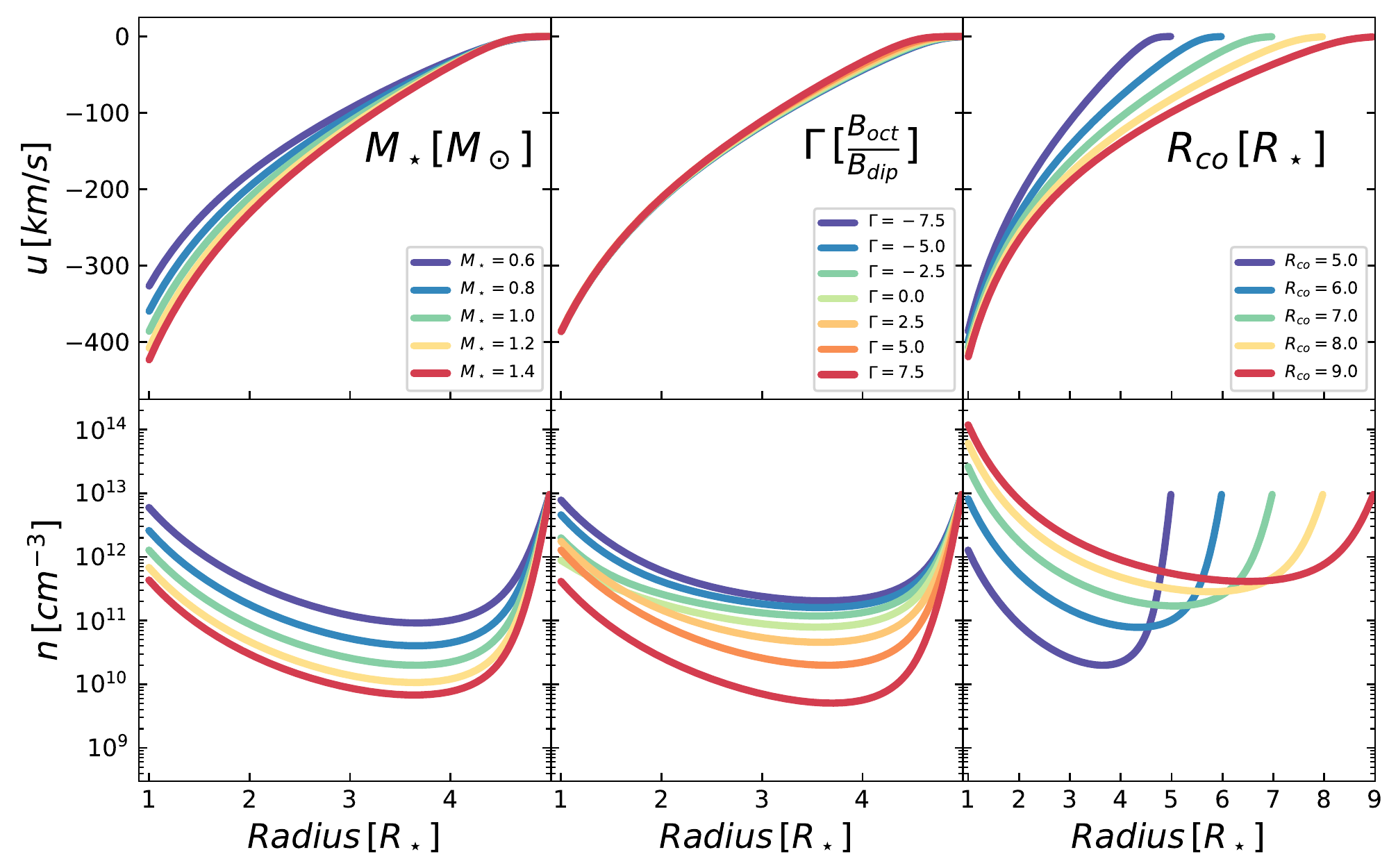}
    \caption{Steady-state solutions for the different values of $M_\star$, $\Gamma$, and $R_{co}$ from the set of models discussed in \S3.3. For the set of parameters we have explored, changes in the kinetic energy flux, $F$, are primarily caused by differences in the density near the star rather than the velocity.}
    \label{fig:steady_state}
\end{figure*}

Next, we consider variants on the base model, where one parameter is changed at a time. To gain insight on the parameters, simulations were first run with simple steady input/output boundary conditions that do not restrict the flow. These simulations converge to steady state and match both the analytic results from \citet{adams12} and the simulations from \citet{robinson17}. The density and velocity solutions are shown in Figure \ref{fig:steady_state}. Next, the steady-state boundary conditions were replaced with the turbulent boundary conditions described in \S 2.2.2. To minimize variation from random effects between models, the random number seed selected for the generation of the turbulent driving function for each model variant is fixed to be the same as the base model shown in  Figures \ref{fig:gettingQ} and \ref{fig:gettingM}. 

\begin{figure*}[h]
    \centering
    \includegraphics[width = 0.9\linewidth]{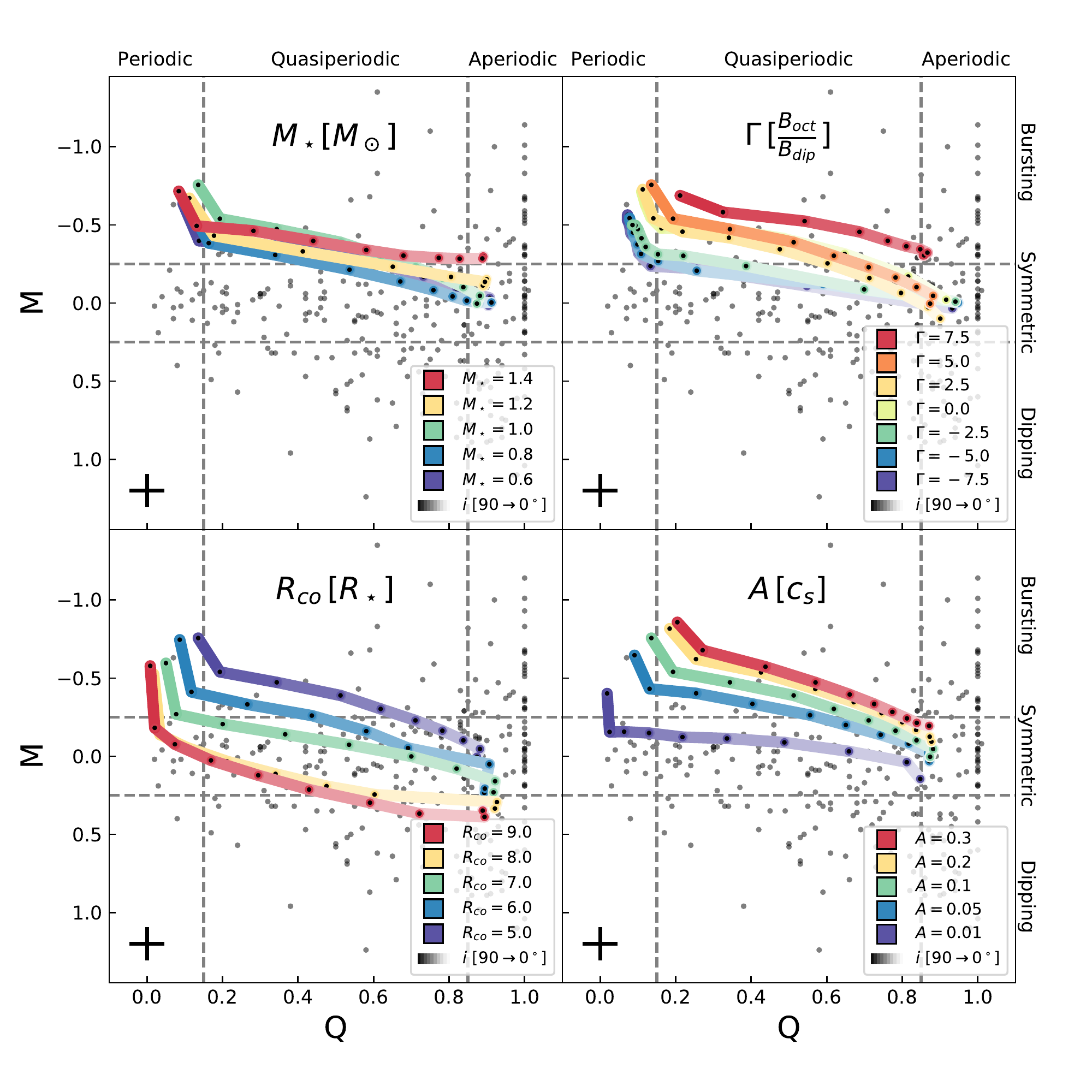}
    \caption{Bivariate plots of metrics $Q$ and $M$ that describe the degree of periodicity and symmetry of the light curves for different parameters. Here, variants on $M_\star$, $\Gamma$, $R_{co}$, and $A$ are shown from a base model of $M_\star = 1.0 \mathrm{M_\odot}$, $\Gamma = 5.0$, $R_{co} = 5.0$ and $A = 0.1$. Individual model realizations are shown as black points along the colored lines. The inclination, $i$, is shown by the darkness of the lines.  The dashed grey lines mark the boundaries between variability classes listed on the top and right axes. An estimate of the model uncertainty is shown by the symbol in the lower left (see \S 3.2). $Q$ and $M$ values from K2 light curves from \citet{cody18} are shown as gray points.}
    \label{fig:QM_variants}
\end{figure*}

\begin{figure*}
    \centering
    \includegraphics[width = 0.98\linewidth]{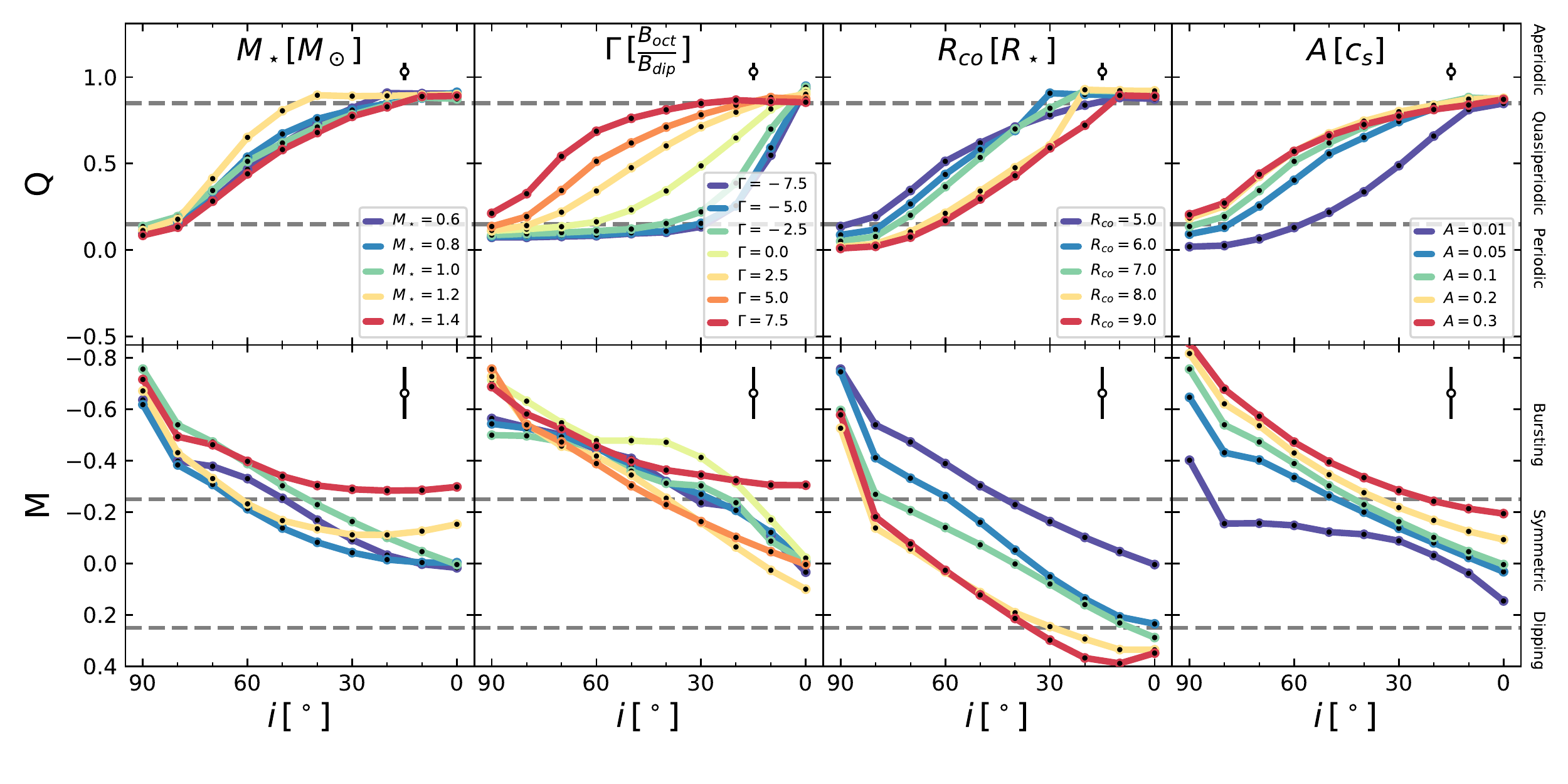}
    \caption{$Q$ and $M$ values as functions of the inclination. Each vertical pair of panels shows model variants of $M_\star$, $\Gamma$, $R_{co}$, and $A$. Individual model realizations are shown by the black points. An estimate of the uncertainty is shown in the upper right, which was derived by changing the seed value of the turbulent driving function for the base model. The dashed gray lines mark the boundaries between the variability classes listed on the far right axis.}
    \label{fig:QMi}
\end{figure*}

\begin{figure*}
    \centering
    \includegraphics[width = 0.98\linewidth]{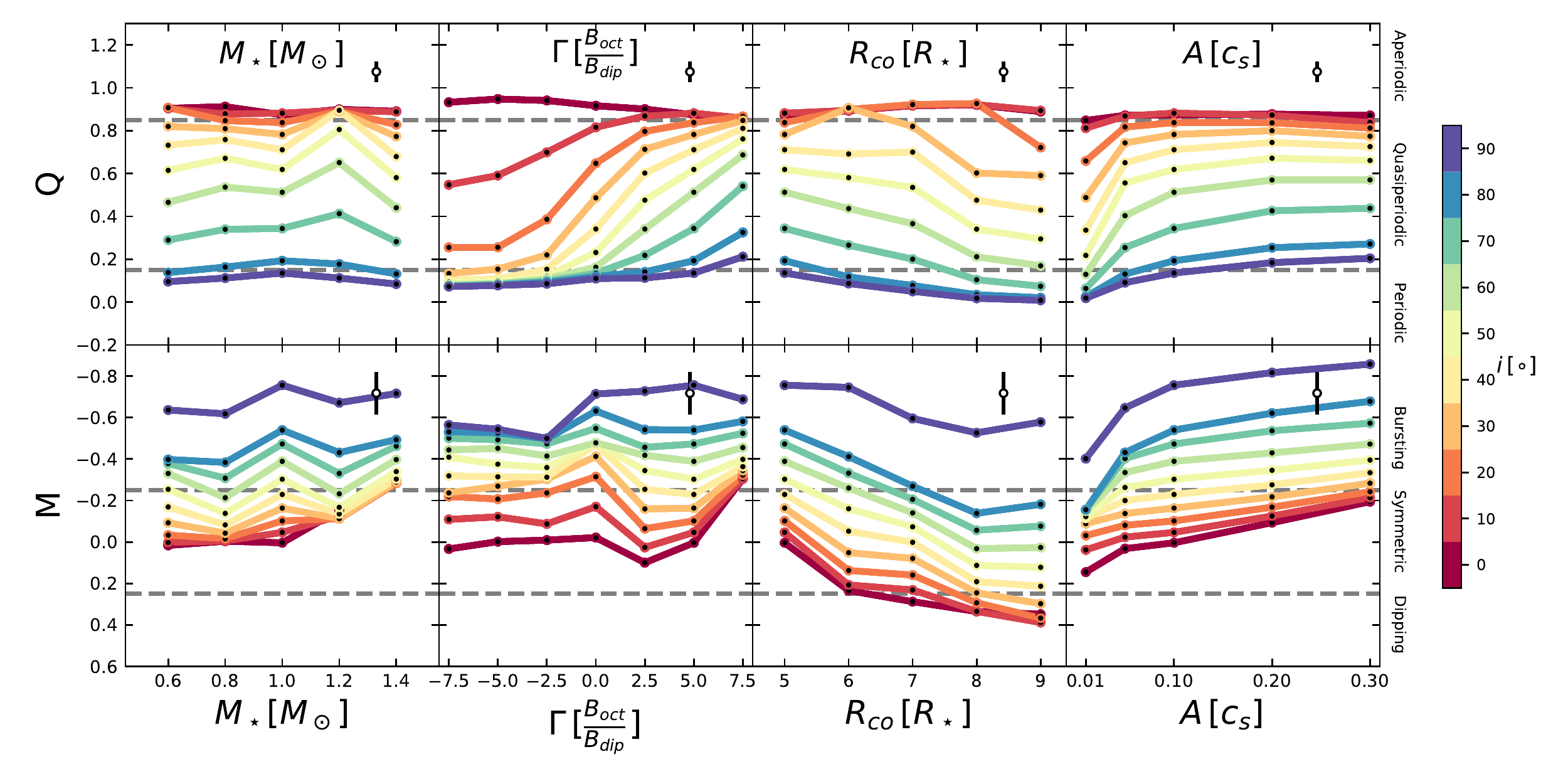}
    \caption{$Q$ and $M$ values as functions of $M_\star$, $\Gamma$, $R_{co}$, and $A$. Individual model realizations are shown by the black points. Darker colors represent models with higher inclinations, with values evenly sampled between $0$ and $90^\circ$. Estimates of the model uncertainties on $Q$ and $M$ are shown by the marker in the upper right corner of each panel. The dashed grey lines mark the boundaries between the variability classes listed on the far right axis.}
    \label{fig:paramplots}
\end{figure*}

In the following sections, the importance of each parameter on variability classification is discussed. The results for all of the models are shown in Figure \ref{fig:QM_variants} displayed in the form of $Q$ vs.\ $M$ bivariate plots. Figure \ref{fig:QMi} shows $Q$ and $M$ as a function of $i$ for the model variants, while Figure \ref{fig:paramplots} demonstrates the effect of changing the other model parameters parameters ($\Gamma$, $R_{co}$, $A$, and $M_\star$). 

\subsubsection{Inclination, $i$}

Under our model construction, $i$ plays  a strong role in determining how periodic the observed signal appears. Objects with low inclinations typically show weaker periodic variations, while objects that are more edge-on show significant rotational modulation. While the inclination at which the model turns from aperiodic to quasiperiodic to periodic depends on the other parameters described above, the trend of an increasing $i$ leading to decreasing $Q$ holds true for all of the models. This is demonstrated in Figure \ref{fig:QMi}. We also find that increasing the inclination tends to lead to light curves that appear more burst dominated. 


\subsubsection{Stellar Mass, $M_\star$}

At a fixed age, increasing the stellar mass also increases the effective temperature and the radius of the star. This changes the depth of the potential well, which changes the flow structure along the accretion column. Under the unitless approach we have taken in the 1D HD simulations, the depth of the potential well is parameterized by the single dimensionless quantity $b \equiv \frac{GM_\star}{R_\star c_s^{2}}$. The effect on $b$ from changing $M_\star$ depends on the age of the star because the radius is set using stellar isochrones \citep{choi16, dotter16}. At an age of 2 Myr, as $M_\star$ increases, $b$ also increases.  Here, we have explored values of $M_\star$ = 0.6, 0.8, 1.0, 1.2, and $1.4 \mathrm{M_\odot}$, which results in $b$ = 511, 618, 711, 794, and 855, respectively.

The first-order effect of increasing $b$ on the flow structure is an increase in the steady-state Mach number of the flow at the stellar surface due to the deeper potential well. However, under our assumption that the outer boundary of the region from which the column draws mass is $R_{co}$, changing the stellar mass also affects rotation. The effect of this on our simulations is parameterized with a dimensionless rotation parameter $\omega \equiv \big(\frac{\Omega R_\star}{c_s}\big)^2$, where $\Omega$ is the rotation rate of the star. Under our assumptions, this can be written as $\omega = \frac{b R_\star^3}{R_{co}^3}$. While increasing $M_\star$ creates a stronger outward centrifugal force in the non-inertial frame, ultimately the effect of increasing gravity is stronger, and the Mach number of the flow increases along the column with increasing $M_\star$. Changing the rotation rate via $M_\star$ in part also controls spot modulation since it sets the rotation period of the star under the constraint of corotation. The range of masses for these models results in rotational periods ranging between 3.3 and 3.5 days. The coordinate system that is used to describe the system is independent of $b$, so changing $M_\star$ does not affect the degree of compression by the magnetic field nor does it change $\gamma_L$ or $\gamma_U$. 

We find that linear relationships between both $Q$ and $M$ and $M_\star$ are not readily apparent. This is best shown by Figure \ref{fig:paramplots}. 
While there may be a slight general trend toward more burst-dominated light curves at higher $M_\star$, particularly at low inclinations, it does not appear to be a monotonic relationship.
The $Q$ metric shows a small peak for the $M_\star = 1.2 \mathrm{M_\odot}$ model at moderate inclinations, but returns to the typical value at $M_\star = 1.4 \mathrm{M_\odot}$. As $M_\star$ increases, the density near the star decreases (see Fig.\ \ref{fig:steady_state}). Interestingly, this leads to a column with a lower kinetic energy flux, since the increased speed of the flow from the larger gravitational potential does not offset the decreased density. This result matches the analytic results presented by \citet{adams12}. Using the same assumptions discussed in \S \ref{ss:basemodel}, we find values of $\dot{M}$ of $2 \times 10^{-10} $ to $6 \times 10^{-10} \, \mathrm{M_\odot \, yr^{-1}}$ (decreasing with $M_\star$) which results in $|B|$ at the poles ranging from 275 to 580 G (also decreasing with $M_\star$).


\subsubsection{Magnetic Field Configuration, $\Gamma$}

The ratio between the dipole and octupole components, $\Gamma$, sets the overarching magnetic field structure of our model. The scale factors of the curvilinear coordinate system set the effective cross-sectional area of the accretion column as a function of radius. These scale factors are a function of $\Gamma$, and as $|\Gamma|$ increases, the flow becomes more constricted near the star \citep[see][for more discussion regarding scale factors]{adams12, robinson17}. Changing $\Gamma$ also changes the shape of the accretion column, which affects components of force vectors along the magnetic field line ($\hat{x} \cdot \hat{p}$ term in the centrifugal force and $\hat{r} \cdot \hat{p}$ term in the gravitational force term).  As shown by Figure \ref{fig:schematic}, increasing $\Gamma$ decreases $\gamma_U$ and $\gamma_L$.

Here, we explore variants with $\Gamma =$ -7.5, -5.0, -2.5, 0.0, 2.5, 5.0, and 7.5. This results in $\gamma_U =$ 58.4, 56.1, 50.2, 26.6, 14.0, 10.4, and $8.6^\circ$, respectively. As $\Gamma$ increases, the steady-state velocity distribution remains relatively unchanged, and the density across most of the column tends to decrease (see Fig.\ \ref{fig:steady_state}). However, near the star, the relation between density and $\Gamma$ is no longer monotonic, and we see a local density maximum around $\Gamma = 2.5$. After implementing the turbulent boundary conditions, we also find a maximum in the $M$ metric near $\Gamma = 2.5$. For values of $\Gamma$ greater than this, we find that as $\Gamma$ increases, $M$ decreases, and the light curve appears more dominated by accretion bursts. 
The pure dipole case shows a more negative value of $M$ than both the $\Gamma = 2.5$ and $5.0$ cases. In the case of $\Gamma = 7.5$, we find that all models appear as bursters. For negative values of $\Gamma$, we found that $M$ was essentially flat with values slightly more positive than that of the pure dipole case at all inclinations. As expected, based on the relationship between $\Gamma$ and $\gamma_L$ and $\gamma_U$, objects smoothly become more aperiodic as $\Gamma$ increases as measured by the $Q$ metric (see the the top panel of Figure \ref{fig:paramplots}). $\dot{M}$ for models with positive, nonzero values of $\Gamma$ under our assumptions range from $7\times 10^{-11}$ to $9\times10^{-10} \, \mathrm{M_\odot \, yr^{-1}}$, and $|B|$ at the poles ranges between $250$ to $360$ G, both decreasing with $\Gamma$. We found that our models with negative values of $\Gamma$ had $\dot{M}$ between $2\times10^{-9}$ to $3\times10^{-9} \mathrm{M_\odot \, yr^{-1}}$, and $|B|$ of $230$ to $1250$ G, both increasing with $|\Gamma|$. For the dipole model, we found $\dot{M} = 1.6\times10^{-9} \, \mathrm{M_\odot \, yr^{-1}}$ and $|B| = 140$ G near the poles.


\subsubsection{Corotation Radius, $R_{co}$}


The field line along which the 1D fluid equations are solved in our model connects the star to disk at the corotation point ($R_{co}$) and the truncation radius is set such that $R_T = 0.5 R_{co}$.
Under our model framework, the choice of $R_{co}$ influences both the rotation rate of the star and which magnetic field lines are traced (see Figure \ref{fig:schematic}). As a consequence, $\gamma_U$ and $\gamma_L$ monotonically decrease with $R_{co}$. Here we have explored inner disk radii of $R_{co} = 5.0$, $6.0$, $7.0$, $8.0$, and $9.0 \, R_\star$. For these values of $R_{co}$, $\gamma_U =$ 10.4, 9.6, 8.9, 8.3, and  $7.8^\circ$, respectively. These values of $R_{co}$ result in rotational periods between 3.3 and 8.1 days.

Changing the inner disk has large consequences on the density distribution along the column in the steady-state solution (see Figure  \ref{fig:steady_state}). The density at the stellar surface spans a full two orders of magnitude between $R_{co} = 5.0 R_\star$ and $R_{co} = 9.0 R_\star$. This is in part because of the decreased effective area of the accretion column near the star in the $R_{co} = 9.0 R_\star$ case due to the associated curvilinear scale factors. However, the additional compression for the field line that crosses the disk at $R_{co}$ is only a factor of about five, which implies that the model is less restrictive to additional mass entering the simulation through the outer boundary as $R_{co}$ increases. This is supported by examining the time-dependent models that show increasing $\dot{M}$ with $R_{co}$, with values ranging between $3 \times 10^{-10}$ and $9\times10^{-9} \, \mathrm{M_\odot \, yr^{-1}}$. Under these values of $\dot{M}$, $|B|$ near the poles ranges between 376 and 5450 G (also increasing with $R_{co}$).

As $R_{co}$ increases, our models tend to be slightly more periodic. This is counter to what the changes in  $\gamma_U$ and $\gamma_L$ would suggest, but the changes are relatively small (see the changes in $\gamma_U$ and $\gamma_L$ from changing $\Gamma$). We found that models with small values of $R_{co}$ showed much stronger bursting behavior for all inclinations. 

\subsubsection{Turbulent Amplitude, $A$}

The turbulent driving function introduces gradients in velocity and density structure near the disk. These gradients lead to the formation of traveling shocks along the accretion column \citep[see][]{robinson17}. The turbulent amplitude, $A$, is the Mach number of the rms velocities of these perturbations at the boundary. Under a \citet{shakura73} $\alpha_\nu$-disk prescription, turbulent perturbations scale as $A = \sqrt{\alpha_\nu}$. This $\alpha_\nu$, not to be confused with the angular spot length, is a dimensionless parameterization of turbulent viscosity. Typical values of $\alpha_\nu$ from modeling protoplanetary disks are on the order of 0.01 \citep[with some scatter; e.g.,][]{hartmann98, andrews09}. Motivated by this, we tested model variants with values of $A = 0.01$, $0.05$, $0.10$, $0.20$, and $0.30$, but caution that $\alpha_\nu$ is generally inferred much further out in the disk than the region that we are simulating.
Recall that the random number seed used to generate the turbulent driving function is fixed across all model variants. 

\begin{figure}
    \centering
    \includegraphics[width = .95\linewidth]{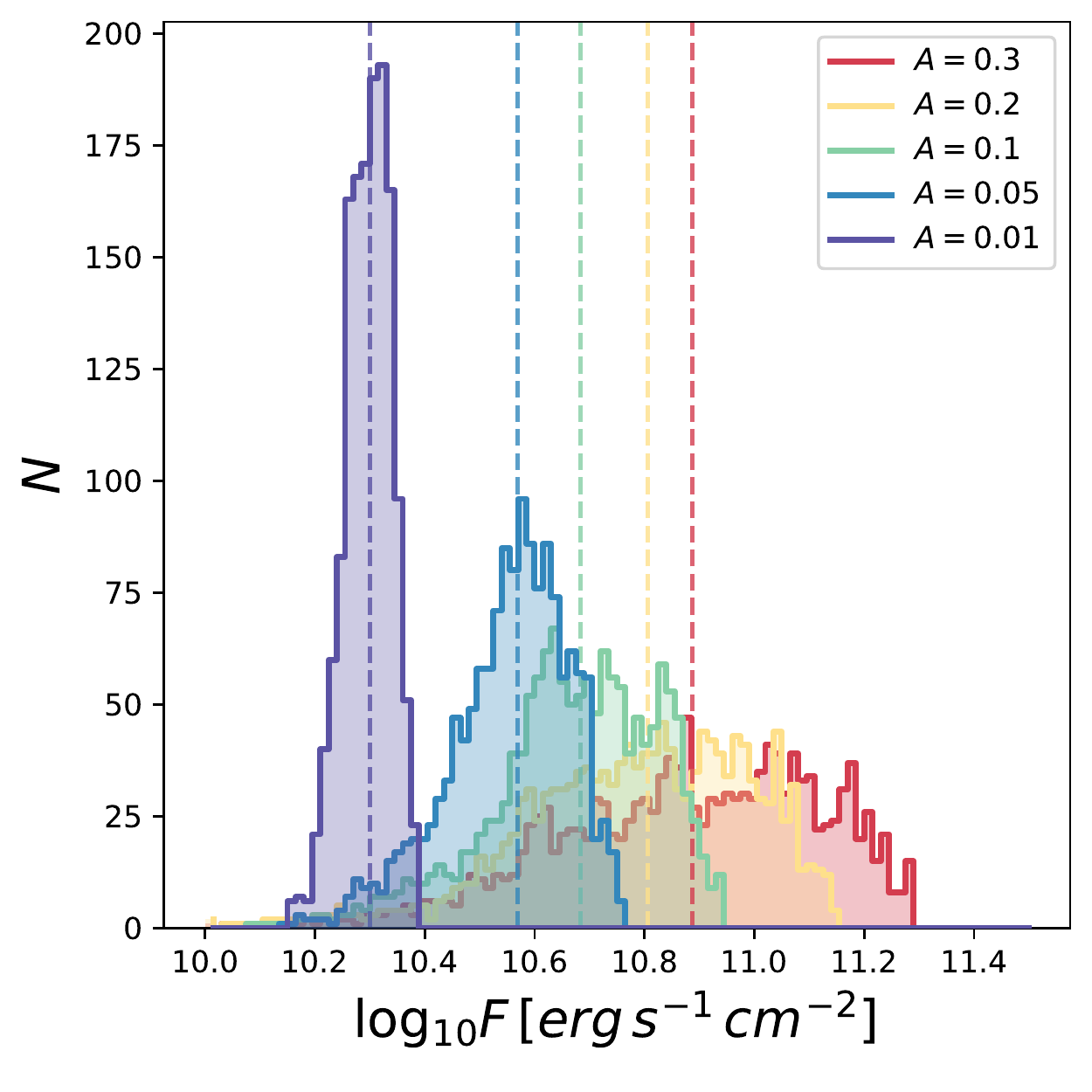}
    \caption{ Histograms showing kinematic energy flux, $F$, at the surface of the star for each time step for different values of turbulent driving amplitude, $A$. The changes in the median value of $F$ (shown by the dashed lines) are nearly entirely due to increased density, as the median velocity at the star changed very little between different values of $A$.}
    \label{fig:A_vs_F}
\end{figure}

As $A$ increases, we find that models monotonically appear more aperiodic as described by the $Q$ metric. Likewise, as $A$ increases, we find that the objects are more dominated by bursts (i.e., $M$ decreases). We find that increasing $A$ also increases the amount of material entering the simulation available for accretion (see Figure \ref{fig:A_vs_F}), which may lead to the some of the differences in the measured $Q$ and $M$ values. Additionally, larger values of $A$ lead to steeper gradients in the outer disk, which aid the formation of the traveling shocks. For this set of models, we found $\dot{M}$ ranging between $1\times10^{-10}$ to $5\times10^{-10} \, \mathrm{M_\odot \, yr^{-1}}$, with $|B|$ between 242 and 475 G near the poles, both increasing with $A$.

\section{Discussion}

We have connected a simple 1D HD simulation of the accretion column \citep[first presented in][] {robinson17} to the accretion shock models of \citet{calvet98, robinson19} and introduced rotational modulation by modeling accretion hot spots as equilatitudinal strips. The output from these simulations was studied using the same tools that have been applied to recent photometric monitoring surveys of accreting young stars in nearby low-mass star-forming regions.  Here we place the results from our simulations in context with these campaigns and discuss implications on variability classifications for different parameters. We also briefly consider two additional complexities that could be explored more in the future: non-rotation-axis-aligned fields and multiple accretion spots.
 
\subsection{Implications for Variability Classes}

Within this work, we have included only mechanisms that can result in variability caused by changes in $\dot{M}$ and rotational effects, so we focus on objects with burster, purely periodic symmetric, quasiperiodic symmetric, and purely stochastic classes. Several of the other classes (quasiperiodic dipper, aperiodic dipper, and eclipsing binary) likely rely on the presence of occulting bodies. Because we do not include any such bodies in our analysis, discussion of these types of objects in general is not included here.  Likewise, the choice of our boundary condition and fixed parameters likely excludes objects that would be classified as long-timescale, unclassifiable, and non-variable.

Under the constraints imposed by our model, we found that inclination plays a large role in determining the periodic behavior of the star as quantified by $Q$. Under the parameters that we explored, we found that typically only objects with low inclinations ($ i \leq 30^\circ$) would be classified as purely stochastic objects. Objects with intermediate inclinations ($ 30^\circ < i < 80^\circ$) tend to fall into the quasiperiodic symmetric classification. Finally, objects that have the highest inclinations ($i \geq 80^\circ$) tend to fall into the category of purely periodic. However, while $i$ is the primary parameter that determines $Q$, the inclination at which the model appears as periodic, quasiperiodic, or aperiodic depends strongly on the other parameters considered here (see Figure \ref{fig:QMi}). For example, the model variant with $\Gamma = 7.5$ is on the border of being classified as stochastic up to an inclination of roughly $40^\circ$. In contrast, the model with $\Gamma = -7.5$ is on the border between quasiperiodic and periodic classifications at the same inclination. Therefore, inclination is likely not the only driver when determining periodic variability class.

We found that the magnetic field geometry (parameterized by $\Gamma$ and $R_{co}$) and the amplitude of the turbulent driving function all had a large impact on $M$, while $M_\star$ did not. We found that models with small $R_{co}$ or large positive values of $\Gamma$ both exhibit bursting behavior with relatively low accretion column densities near the star (see Figure \ref{fig:steady_state}). In contrast, models with large values of $A$ also show bursts with increasing column density (Figure \ref{fig:A_vs_F}). Contemporaneous NUV - NIR spectra can be used to break degeneracies between accretion column density and surface coverage \citep[e.g.,][]{ingleby13, ingleby15, robinson19}, and the magnetic field can be directly probed with spectropolarimetric observations, \citep[e.g.,][]{donati10a}. Combined, these techniques may be able to distinguish between these scenarios. 
With regards to $M_\star$, \citet{cody18} did not find an obvious relationship between variability classification and $M_\star$ using K2 observations and spectral types as proxies for stellar masses. This is in agreement with the non-monotonic relationships we found between $M_\star$ and $Q$ and $M$.

While the range of models that we explored does not cover all of parameter space, we can speculate on some basic generalizations about variability class populations given that our selection of base model parameters is somewhat representative of a typical CTTS.  We found that most of our models fell into the quasiperiodic symmetric classification, suggesting that these types of light curves might be common. This is in agreement with previous photometric monitoring campaigns that also showed the largest fraction of stars in the quasiperiodic symmetric class \citep[$17\pm3 \%$ to $29 \pm 3\%$, depending on the star-forming region;][]{cody18}. 
We also found that under the range of parameters that we explored, this model can produce light curves that fall into all of the classifications associated with changes in $\dot{M}$. This suggests that the some of the observed variability could feasibly be driven by Kolmogorav-style turbulence in the inner disk. However, we caution further conjecture at this point regarding the entire population because the variability classification of the model variants strongly depends on the base model parameters. Additionally, observational biases that affect studies of variability in young stars are not considered here (e.g., the star is unlikely to be observable at high inclinations due to disk obscuration). These models also lack the ability to reproduce fluid instabilities that 3D simulations can reveal. Additionally, only a single hot spot has been considered (see \S \ref{ss:multiplespots}).
 
\subsection{Nonaligned Magnetic Field Structure}

One of the largest assumptions in our model is the presence of a perfectly rotation-axis-aligned magnetic field that can be described using a multipole decomposition. Zeeman-Doppler imaging of the magnetic field structure of young stars has shown that the magnetic fields are not necessarily perfectly aligned with the rotation axis \citep{donati07, donati10b}. This misalignment, along with any possible toroidal components, causes the accretion hot spot geometry to become significantly more complex. 

We can begin to consider the effects of such a misaligned field on our model construction. If deviations from a purely rotation-axis-aligned magnetic field are included, the easiest effect to consider without detailed HD simulations is the change in $\gamma_U$ and $\gamma_L$.  Models with larger values of $\gamma_U$ and $\gamma_L$ tend to be more periodic since the effect of rotational modulation of the light curve is enhanced. With the introduction of a misalignment, the magnetic field connecting the disk to the star becomes a function of $\phi$ in the corotating frame. If magnetic field lines further from the axis of rotation are favored for mass loading because the gravitational force vector is more closely aligned with the magnetic field line at the magnetic footprint at the inner edge of the disk \citep[predicted by 3D MHD simulations, e.g., ][]{romanova03, romanova04}, then misalignments would tend to increase the periodicity of the observed light curve. However, as demonstrated in this work, changing the magnetic field geometry can drastically change the flow along the accretion column, thus making more detailed modeling required for more concrete predictions.

\subsection{Multiple Spots \label{ss:multiplespots}}

The mass deposited onto the star in our models is uniformly spread across a single hot spot. Modeling of NUV spectra, which can break the degeneracy in Figure \ref{fig:U-V}, has shown that multiple accretion column densities are typically necessary to fully reproduce observations within single observations \citep{ingleby13, ingleby15, robinson19}. This suggests either that multiple accretion hot spots exist with different densities and/or that gradients occur across the  single spot. Adding more spots introduces several new free parameters that were not necessary to consider in the single-column case. In this work, we also did not consider the conjugate spot on the opposite pole. This spot may be visible for specific models at higher inclinations (e.g., models with negative values of $\Gamma$).

Using two of the base model realizations with different random seeds (see \S 3.2), we performed exploratory tests to examine the effect of adding a second nonoverlapping spot. The phase offset and the ratio between spot areas was varied while the total area covered was fixed to be the same as a single spot with $\alpha = 90^\circ$. We also fixed $i = 60^\circ$ for simplicity. $Q$ and $M$ were then measured for each phase-ratio pair. We explored five spot area ratio cases: $5:1$, $3:1$, $1:1$, $1:3$, and $1:5$ and five evenly sampled phase offsets between $120^\circ$ and $240^\circ$ for each ratio. We found limited impact on $Q$ and $M$ between different phase offsets of the spots. The mean standard deviation for $Q$ and $M$ when fixing the spot size ratio and varying the phase offset was 0.04 and 0.07, respectively. We found that changing the ratio of spot sizes has a larger effect on $Q$ and $M$. The mean standard deviations of $Q$ and $M$ between models with a fixed phase offset and a varied spot ratio were 0.15 and 0.14, respectively. The mean $Q$ and $M$ values for the two-column realizations were 0.77, and  -0.35, which are similar to the the two single-column models with values of 0.51 and 0.72 for $Q$ and -0.38 and -0.43 for $M$. However, trends between the spot area ratio and the change in $Q$ and $M$ are not easily identified. Even light curves of conjugate pairs of spot size ratios (e.g., $3:1$ and $1:3$) with the same phase offset can appear remarkably different. This reflects the dependence on the stochastic nature of the chosen driving functions, and the spread in the metrics is similar to that calculated between single-column models with different random seeds \S 3.2. Ultimately, adding additional spots, via either multiple accretion streams or conjugate spots on the opposite hemisphere, tends to act as a contributing source of uncertainty that has not been considered in the single-column analysis presented in this work and merits further exploration.

\section{Summary}
We created synthetic light curves of accreting pre-main-sequence stars using a set of 1D HD simulations of the magnetosphere that are coupled with an nLTE accretion shock model and a simple model of rotational modulation. Our models were constructed under the assumption of a dipole or a dipole + octupole rotation-axis-aligned magnetic field. Perturbations were introduced to steady-state flow solutions using boundary conditions that aimed to mimic turbulence in the inner disk. Specifically, we tested the effect of the stellar mass, inclination, magnetic field geometry, the inner disk radius, and the level of turbulence in the inner disk.

We then used the previously developed metrics $Q$ and $M$ to measure the light curve periodicity and symmetry around the mean, respectively, in order to facilitate comparisons between patterns present in our model variants and the results from month-long photometric monitoring campaigns of young stars \citep[e.g., ][]{cody14, cody18}. We focus on variability classes that can, in part, be explained by changes in $\dot{M}$. These include bursters, purely periodic symmetric, quasiperiodic symmetric, and purely stochastic classes. We found that Kolmogorav turbulence in the inner disk can drive a wide degree of variability behavior associated with these classes. Our major findings are summarized as follows. 

\begin{enumerate}

    \item Clear monotonic trends between $M_\star$ and the $Q$ and $M$ metrics are difficult to identify, which is in agreement with previous observations of CTTS. 
    
    \item We found that the periodic behavior of the light curve as measured by $Q$ is primarily controlled by the inclination, $i$. Typically, we find that objects with $i \leq 30^\circ$ appear as purely stochastic, $30^\circ < i < 80^\circ$ as quasiperiodic, and $i \geq 80^\circ$ as purely periodic. However, the value of $i$ where the object appears periodic, quasiperiodic, or aperiodic depends to some degree on the magnetic field geometry, the inner disk radius, and the level of turbulence in the inner disk. 
    
    \item We found that the symmetric behavior of the light curve as measured by $M$ is also a function of the inclination, magnetic field composition, inner disk radius, and turbulent amplitude. Objects with either a pure dipole field, large aligned octupole components, or a small $R_{co}$ tend to exhibit more burst-dominated behavior. Objects with anti-aligned octupole components tend to show slightly less burst-dominated light curves than objects with pure dipolar fields at identical inclinations. Increasing the amplitude of the turbulent perturbations also led to more bursts.

    \item Our models suggest that it may be possible to separate driving parameters behind variability observed in light curves using NUV spectra to measure the kinetic energy flux of the accretion flow and spectropolarimetric measurements of the magnetic field.
    
    \item Complexities such as multiple columns and non-axis-aligned magnetic fields can modify the measured $Q$ and $M$ parameters. We suggest that a non-axis-aligned field may lead to more periodic light curves. Our initial tests found that when additional columns are included, $Q$ remains similar to the single column model while $M$ varies widely as a function of the ratio between spot sizes. We note that more work is necessary to make further predictions for either additional complexity.
    
    \item We suggest a slight adjustment to the algorithm for measuring rotation periods when measuring $Q$. We found that weighting the Lomb--Scargle periodogram with a Gaussian whose parameters are derived from the ACF gave a more accurate estimate than previous methods for our synthetic light curves (whose true period is known a priori).
    
\end{enumerate}

More continuous light curves of well-studied CTTS will become available from TESS in the upcoming years. This work attempts to establish the groundwork for interpretation of those data with a relatively simple approach. 

C.E.R. and C.C.E. acknowledge support from the National Science Foundation under CAREER grant No. AST-1455042 and the Sloan Foundation.
The authors also acknowledge funding support from NASA grant No. 80NSSC19K1712. 
J.E.O. is supported by a Royal Society Univiersity Research Fellowship. The authors thank the referee, S. Gregory, whose insightful comments have greatly improved the quality of this work. 

\bibliography{biblio}

\end{document}